\documentclass[twocolumn]{revtex4}

\usepackage{graphicx}
\usepackage{dcolumn}
\usepackage{amsmath}

\begin{document}

\title{Vortices and chirality in multi-band
superconductors
}

\author{Takashi Yanagisawa$^{a,b}$,Yasumoto Tanaka$^{a}$, Izumi Hase$^{a,b}$
and Kunihiko Yamaji$^{a,b}$
}

\affiliation{$^a$Electronics and Photonics Research
Institute, National Institute of Advanced Industrial Science and Technology (AIST),
Tsukuba Central 2, 1-1-1 Umezono, Tsukuba 305-8568, Japan\\
$^b$CREST, Japan Science and Technology Agency (JST), Kawaguchi-shi, 
Saitama 332-0012, Japan\\
}

%\kword{multi-band superconductor, chirality, vortex, time-reversal symmetry
%breaking}

\begin{abstract}
We investigate some significant properties of multi-band superconductors.
They are time-reversal symmetry breaking, chirality and fractional quantum
flux vortices in three-band superconductors.
The BCS (Bardeen-Cooper-Schrieffer) gap equation has a solution with time-reversal
symmetry breaking in some cases.
We derive the Ginzburg-Landau free energy from the BCS microscopic theory.
The frustrating pairing interaction among Fermi surfaces leads to a state with broken
time-reversal symmetry, that is, a chiral solution.
The Ginzburg-Landau equation for three-component superconductors 
leads to a double sine-Gordon model. 
A kink solution exists to this equation as in the conventional sine-Gordon model.
In the chiral region of the double sine-Gordon model, an inequality of Bogomol'nyi 
type holds, 
and fractional-$\pi$ kink solutions exist with the topological charge $Q$.
This yields multi-vortex bound states in three-band superconductors. 
\end{abstract}

\maketitle

\section{Introduction}

Since the discovery of oxypnictides LaFeAsO$_{1-x}$F$_x$\cite{kam08},
BaFe$_2$As$_2$\cite{rot08}, LiFeAs\cite{pit08,tap08} and 
Fe$_{1+x}$Se\cite{hsu08,miz08},  
the Fe pnictides high-temperature superconductors have attracted extensive
attention.
There are numerous experimental studies regarding the electronic states of
the new family of iron-based 
superconductor\cite{che08a,che08b,cru08,nak08,sha08,lue08}.
The undoped samples exhibit the antiferromagnetic transition\cite{cru08,nak08},
and show the superconducting transition with electron doping\cite{kam08}.
The band structure calculations indicate that the Fermi surfaces are composed
of two hole-like cylinders around $\Gamma$, a three-dimensional Fermi surface,
and two electron-like cylinder around M for LaFeAsO\cite{sin08}.
This family of iron pnictides is characterized by multi Fermi surfaces, and
theoretical studies have been based on multi-band models with electronic
interactions\cite{kur08,ike08,ohn08}. 
An importance of multi-band structure is obviously exhibited in  recent
measurements of the Fe isotope effect\cite{liu08,shi09}.
The inverse isotope effect in (Ba,K)Fe$_2$As$_2$ can be understood by the multi-band
model with competing inter-band interactions\cite{yan09}.
The two-gap theory of superconductivity has a long history, and is the generalization
of the BCS theory to the case with two conduction bands\cite{suh59,kon63,leg66}

The objective of this paper is to study time-reversal symmetry breaking and
fractional quantum-flux vortices in multi-component superconductors. 
We investigate physical properties of the superconducting state with
time-reversal symmetry breaking.  
The $3\times 3$ gap equation has such a solution if the pairing
interactions satisfy some conditions.
To investigate vortex solutions,
we derive the Ginzburg-Landau functional\cite{gei67,gur03,zhi04} 
from the microscopic theory
of multi-component superconductivity.
The importance of phase dynamics has been pointed out 
previously\cite{leg66,izu90,tan01,tan02,bab02,leg04}.
The phase variables of the order parameters may lead to a new state as a minimum of
the Ginzburg-Landau potential.  If the phase of the gap function takes a fractional
value other than 0 or $\pi$, a significant state, called the chiral 
state\cite{tan10a,tan10b},
appears and the time reversal symmetry is broken.
This unconventional state is induced from the Josephson terms with frustrating
pairing interactions.  The signs of the Josephson terms play an important role in
determining the ground state.

The three-band model leads to the double sine-Gordon model as a model to describe 
the dynamics of phase variables.  This model is regarded as a generalization of the
usual sine-Gordon model\cite{raja} and  also has kink solutions as in the
sine-Gordon model.  A new feature that appears first in the double sine-Gordon
model is that we have fractional-$\pi$ kink solutions in the chiral.
We can define the topological charge from
the topological conserved current.
In the chiral region, fractional flux vortices may exist on a domain wall of the
kink. 

The paper is organized as follows.  
In Section II, we present the model Hamiltonian that is considered in this paper.
In Section III, we show the model that exhibits time-reversal symmetry breaking.
In Section IV, the Ginzburg-Landau functional
for three-band superconductors is derived by using the Gor'kov method.
Each term of this functional is expressed in terms of the matrix $G=(g_{ij})$
where $g_{ij}$ is the pairing interaction between the bands $i$ and $j$.
Here we investigate the magnetic properties near the upper critical field.
In Section V we consider the phase dynamics of the order parameters.
We investigate the ground state of the potential composed of phase variables
and show the existence of the chiral region with time reversal symmetry breaking.
The double sine-Gordon model appears in the three-band model and exhibits
unique properties that are not contained in the conventional sine-Gordon model.
In Section VI we discuss the existence of fractional flux vortices and
multi-vortex bound states. 
We also discuss applications to multi-band superconductors such as Fe pnictides.

\section{Three-Band BCS Model}

We adopt the three-band BCS model with the attractive interactions
\begin{eqnarray}
H&=& \sum_{i\sigma}\int d{\bf r}\psi_{i\sigma}^{\dag}({\bf r})K_i({\bf r})
\psi_{i\sigma}({\bf r}) \nonumber\\
&-& \sum_{ij}g_{ij}\int d{\bf r}\psi_{i\uparrow}^{\dag}({\bf r})
\psi_{i\downarrow}^{\dag}({\bf r})\psi_{j\downarrow}({\bf r})
\psi_{j\uparrow}({\bf r}),\nonumber\\
\end{eqnarray}
where $i$ and $j$ (=1,2,3)  are band indices.
$K_i({\bf r})$ stands for the kinetic operator.
We assume that $g_{ij}=g_{ji}^*$.
The second term is the pairing interaction and $g_{ij}$ are coupling constants.
The mean-field Hamiltonian is
\begin{eqnarray}
H_{MF}&=& \sum_{i}\int d{\bf r}\Big[\sum_{\sigma}\psi_{i\sigma}^{\dag}({\bf r})
K_i({\bf r})\psi_{i\sigma}({\bf r}) \nonumber\\
&+&\Delta_i({\bf r})\psi_{i\uparrow}^{\dag}({\bf r})\psi_{i\downarrow}^{\dag}({\bf r})
+\Delta_i^*({\bf r})\psi_{i\downarrow}({\bf r})\psi_{i\uparrow}({\bf r})\Big],
\end{eqnarray}
where the gap function in each band is defined by
\begin{equation}
\Delta_i({\bf r})= -\sum_j g_{ij}\langle\psi_{j\downarrow}({\bf r})\psi_{j\uparrow}
({\bf r})\rangle,
\end{equation}
and its complex conjugate is
\begin{equation}
\Delta_i^*({\bf r})= -\sum_j g_{ji}\langle\psi_{j\uparrow}^{\dag}({\bf r})
\psi_{j\downarrow}^{\dag}({\bf r})\rangle.
\end{equation}
We define Green's functions as follows\cite{gor59},
\begin{equation}
G_{j\sigma\sigma'}(x-x')= -\langle T_{\tau}\psi_{j\sigma}(x)\psi_{j\sigma'}^{\dag}
(x')\rangle,
\end{equation}
\begin{equation}
F_{j\sigma\sigma'}^+ (x-x')= \langle T_{\tau}\psi_{j\sigma}^{\dag}(x)
\psi_{j\sigma'}^{\dag}(x')\rangle,
\end{equation}
where $T_{\tau}$ is the time-ordering operator and we use the notation $x=(\tau,{\bf r})$.
In terms of the Green's functions, the gap functions satisfy the system of equations
\begin{eqnarray}
\Delta_i^*({\bf r})&=&\sum_{j}g_{ij}^*F_{j\downarrow\uparrow}^+(\tau'=\tau+0; {\bf r},{\bf r})
\nonumber\\
&=& \sum_j g_{ij}^*\frac{1}{\beta}\sum_n F_{j\downarrow\uparrow}^+
(i\omega_n;{\bf r},{\bf r}).
\label{gapeq}
\end{eqnarray}
This yields the gap equation,
\begin{equation}
\Delta_i= \sum_j g_{ij}N_j\Delta_j \int d\xi_j\frac{1}{E_j}{\rm tanh}\left(
\frac{E_j}{2T}\right),
\end{equation}
where $E_j=\sqrt{\xi_j^2+|\Delta_j|^2}$ and $T$ is the temperature where we set
Boltzmann constant $k_B$ to unity.
$N_j$ is the density of states at the Fermi surface.
Since all the bands couple with each other through mutual interactions $g_{ij}$,
we have one critical temperature $T_c$\cite{suh59}.
At the critical temperature $T=T_c$, this equation reads\cite{gei67}
\begin{equation}
\Delta_i= {\rm ln}\left(\frac{2e^{\gamma}\omega_c}{\pi T_c}\right)\sum_j
g_{ij}N_j\Delta_j,
\label{gaplin}
\end{equation}
for the cutoff energy $\omega_c$.  $\gamma$ denotes the Euler constant.  
Here we assume the same cutoff energy in
all the interactions.
The critical temperature $T_c$ is obtained as a solution to the equation
\begin{eqnarray}
0=\left|
\begin{array}{ccc}
\eta g_{11}N_1-1  &  \eta g_{12}N_2  &  \eta g_{13}N_3  \\ 
\eta g_{21}N_1   & \eta g_{22}N_2-1  &  \eta g_{23}N_3  \\ 
\eta g_{31}N_1   &  \eta g_{32}N_2   &  \eta g_{33}N_3-1  \\ 
\end{array}
\right|,
\label{determ}
\end{eqnarray}
where $\eta={\rm ln}(2e^{\gamma}\omega_c/\pi T_c)$.
The system of equations in eq.(\ref{gapeq}) yields a set of differential
equations\cite{gei67}
\begin{eqnarray}
\Delta_j^*({\bf r})&=& {\rm ln}\left(\frac{2e^{\gamma}\omega_c}{\pi T}\right)
\sum_{\ell}g_{j\ell}^*N_{\ell}\Delta_{\ell}^*({\bf r})\nonumber\\
&+& \frac{7\zeta(3)}{48(\pi T_c)^2}\sum_{\ell}g_{j\ell}^*N_{\ell}v_{\ell}^2
\left(\nabla+i\frac{2e}{\hbar c}{\bf A}\right)^2\Delta_{\ell}^*({\bf r})\nonumber\\
&-& \frac{7\zeta(3)}{8(\pi T_c)^2}\sum_{\ell}g_{j\ell}^*N_{\ell}\Delta_{\ell}^*
({\bf r})|\Delta_{\ell}({\bf r})|^2.
\end{eqnarray}
Here, $e$ is the charge of the electron, and
$v_{\ell}$ is the electron velocity at the Fermi surface in the $\ell$-th band.
The Planck constant $\hbar$ has been dropped except in front of the vector
potential ${\bf A}$.
We set $a_{mn}= g_{mn}N_n$.  Then, the equation for $\Delta_j$ is
\begin{equation}
0=\left[ a_{jj}{\rm ln}\left(\frac{2e^{\gamma}\omega_c}{\pi T}\right)-1\right]
\Delta_j
+{\rm ln}\left(\frac{2e^{\gamma}\omega_c}{\pi T}\right)\sum_{\ell(\neq j)}
a_{j\ell}\Delta_{\ell}+\cdots.
\end{equation}
In the second term of the right-hand side we can replace $T$ by $T_c$ near the
transition temperature.
From eq.(\ref{gaplin}) we obtain
\begin{eqnarray}
\eta\left(
\begin{array}{c}
\Delta_1 \\
\Delta_2 \\
\Delta_3 \\
\end{array}
\right)=
\left(
\begin{array}{ccc}
b_{11}  &  b_{12}  &  b_{13}  \\
b_{21}  &  b_{22}  &  b_{23}  \\
b_{31}  &  b_{32}  &  b_{33}  \\
\end{array}
\right)
\left(
\begin{array}{c}
\Delta_1 \\
\Delta_2 \\
\Delta_3 \\
\end{array}
\right),
\end{eqnarray}
where $b_{j\ell}$ are elements of the inverse of the matrix $A=(a_{ij})$:
$A^{-1}=(b_{j\ell})$.
Then the equation for $\Delta_j$ reads
\begin{equation}
0= \left[ a_{jj}{\rm ln}\left(\frac{2e^{\gamma}\omega_c}{\pi T}\right)-1\right]
\Delta_j
+ \sum_{\ell(\neq j)}
a_{j\ell}\sum_{m}b_{\ell m}\Delta_m+\cdots.
\label{eqj}
\end{equation}
For example, for $j=1$ we obtain
\begin{eqnarray}
0&=& g_{11}\Big[\left( N_1\ln\left(\frac{2e^{\gamma}\omega_c}{\pi T}\right)
-\frac{1}{\det G}(G^{-1})_{11}\right)\Delta_1 \nonumber\\
&-& \frac{1}{\det G}(G^{-1})_{12}\Delta_2
-\frac{1}{\det G}(G^{-1})_{13}\Delta_3 \Big]+\cdots,
\end{eqnarray}
where $G=(g_{ij})$ is the matrix of coupling constants.
To obtain the multi-band Ginzburg-Landau functional, we multiply eq.(\ref{eqj})
by $\Delta_j^*N_j$ and take a summation with respect to $j$.
We use the gap equation $\Delta_{\ell}^*=\eta \sum_jg_{\ell j}^*N_j\Delta_j^*$
at $T=T_c$.  Then the energy functional density $f$ is
\begin{eqnarray}
f&=& -\sum_j \frac{g_{jj}N_j}{\langle gN\rangle}
\left(N_j\ln\frac{2e^{\gamma}\omega_c}{\pi T}-(G^{-1})_{jj}\right)
|\Delta_j|^2 \nonumber\\
&+& \sum_{j\ell}\frac{g_{jj}N_j}{\langle gN\rangle}\Delta_j^*(G^{-1})_{j\ell}
\Delta_{\ell} \nonumber\\
&-& \frac{7\zeta(3)}{48\pi^2T_c^2}\sum_{\ell}N_{\ell}v_{\ell}^2
\Delta_{\ell}^*\left(\nabla-i\frac{2e}{\hbar c}{\bf A}\right)^2\Delta_{\ell}
\nonumber\\
&+& \frac{7\zeta(3)}{16\pi^2T_c^2}\sum_{\ell}N_{\ell}|\Delta_{\ell}|^4.
\end{eqnarray}
Here we set $\eta=1/\langle gN\rangle$, and
we neglect unimportant constants $g_{jj}N_j/\langle gN\rangle$
in the following.
The fourth order term is simply given by $(|\Delta_{\ell}|^2)^2$.
Please note that out functional is relevant near $T_c$.

\section{Time-Reversal Symmetry Breaking}

The three-band superconductor exhibits a significant state with broken
time-reversal symmetry, that is not shown in single-band and two-band
superconductors.
This originates from the fact that the gap equation for the three-band model
has complex eigen-functions in some cases.
The gap equation is
\begin{eqnarray}
\lambda\left(
\begin{array}{c}
\Delta_1 \\
\Delta_2 \\
\Delta_3 \\
\end{array}
\right)=
\left(
\begin{array}{ccc}
g_{11}N_1  &  g_{12}N_2  &  g_{13}N_3  \\
g_{21}N_1  &  g_{22}N_2  &  g_{23}N_3  \\
g_{31}N_1  &  g_{32}N_2  &  g_{33}N_3  \\
\end{array}
\right)
\left(
\begin{array}{c}
\Delta_1 \\
\Delta_2 \\
\Delta_3 \\
\end{array}
\right),
\end{eqnarray}
where $1/\lambda=\ln(2e^{\gamma}\omega_c/\pi T_c)$.
We assume that the matrix $(g_{ij}N_j)$ is real and symmetric.
When two eigenvalues are degenerate, we obtain two complex eigenvectors
by making linear combinations.
For example, let us consider a matrix $A=(A_{ij})$ given by
\begin{eqnarray}
A_{11}&=& J(\cos^2\theta\cos^2\varphi+\sin^2\varphi)+D  \\  
A_{12}&=& A_{21}= -J\sin^2\theta\sin\varphi\cos\varphi  \\
A_{13}&=& A_{31}= -J\sin\theta\cos\theta\cos\varphi \\
A_{22}&=& J(\cos^2\theta\sin^2\varphi+\cos^2\varphi)+D  \\
A_{23}&=& A_{32}= -J\sin\theta\cos\theta\sin\varphi \\
A_{33}&=& J\sin^2\theta+D  \\
\label{ematrix}
\end{eqnarray}
where $J$ is a positive constant and $D$ is also a constant satisfying $J+D>0$.  
$J$ indicates the strength of interband coupling constants.
The degenerate eigenvalue is given by
\begin{equation}
\lambda= J+D.
\end{equation}
$\theta$ and $\varphi$ are Euler angles in the range of
$0\leq\theta\leq\pi$ and $0\leq\varphi\leq 2\pi$.

The eigenvectors of the matrix in eq.(\ref{ematrix}) that correspond to the
eigenvalue $\lambda=J+D$ are
\begin{eqnarray}
\left(
\begin{array}{c}
\cos\theta\cos\varphi \\
\cos\theta\sin\varphi \\
-\sin\theta \\
\end{array}
\right),
\left(
\begin{array}{c}
\sin\varphi \\
-\cos\varphi \\
0 \\
\end{array}
\right).
\end{eqnarray}
When $\varphi=\pi/4$, they are
\begin{eqnarray}
\left(
\begin{array}{c}
e^{i\phi} \\
e^{-i\phi} \\
1 \\
\end{array}
\right),
\left(
\begin{array}{c}
e^{-i\phi} \\
e^{i\phi} \\
1 \\
\end{array}
\right),
\end{eqnarray}
where the angle $\phi$ satisfies $\cos\phi=-1/(\sqrt{2}\tan\theta)$, and
we assume that $|\sqrt{2}\tan\theta|\geq 1$.
In this way the time-reversal symmetry broken state is obtained from the
gap equation.
If we set $\varphi=\pi/4$ and $\tan\theta=\sqrt{2}$, we obtain the case
with three equivalent interband pairing interactions, that is, the three
equivalent bands\cite{sta10}.  In this case we have $\phi=2\pi/3$, and thus the phase
difference between order parameters of three bands is $2\pi/3$.

\section{Three-Gap Ginzburg-Landau Functional}

In this section, we derive the Ginzburg-Landau free energy from the BCS model
to investigate chirality and kinks.
From the results in the section I\hspace{-.1em}I,
the Ginzburg-Landau functional for three-gap superconductors has the form
\begin{eqnarray}
F&=& \int d{\bf r}\Biggl[ \sum_{j=1}^3 \alpha_j|\psi_j|^2
+\frac{1}{2}\sum_{j=1}^3 \beta_j|\psi_j|^4  \nonumber\\
&-& (\gamma_{12}\psi_1^*\psi_2
+ \gamma_{21}\psi_2^*\psi_1)\nonumber\\
&-& (\gamma_{23}\psi_2^*\psi_3+\gamma_{32}\psi_3^*\psi_2)
-(\gamma_{31}\psi_3^*\psi_1+\gamma_{13}\psi_1^*\psi_3) \nonumber\\
&+& \sum_{j=1}^3K_{j} \Bigl|\left(\nabla+i\frac{2\pi}{\phi_0}{\bf A}\right)\psi_j \Bigr|^2 
+\frac{1}{8\pi}{\bf H}^2 \Biggr].
\end{eqnarray}
$\beta_j$ are constants and $\phi_0$ is the flux quantum,
\begin{equation}
\phi_0= \frac{hc}{|e^*|}=\frac{hc}{2|e|}.
\end{equation}
The coefficients of bilinear terms are
\begin{equation}
\alpha_j 
= -\left[N_j{\rm ln}\left(\frac{2e^{\gamma}\omega_c}{\pi T}\right)
-(G^{-1})_{jj}\right],
\end{equation}
\begin{equation}
\gamma_{ij}= -(G^{-1})_{ij},
\end{equation}
for $i,j=1,2,\cdots$.
The formula for more than three bands is straightforwardly obtained in
terms of $G^{-1}$. 
The critical temperature is given by, with $\eta$ in eq.(\ref{determ}),
\begin{equation}
T_c= \frac{2e^{\gamma}\omega_c}{\pi}e^{-\eta}.
\end{equation}
For the two-band case, if we assume that gap functions are constant, 
$\eta$ is explicitly written as
\begin{eqnarray}
\eta&=& \frac{1}{2{\rm det}A}\left[a_{11}+a_{22}-\sqrt{(a_{11}-a_{22})^2
+4a_{12}a_{21}}\right]\nonumber\\
&=& \frac{g_{11}/N_2+g_{22}/N_1}{2{\rm det}G} \nonumber\\
&-&\frac{1}{{\rm det}G}\sqrt{ \frac{1}{4}\left(\frac{g_{11}}{N_2}-\frac{g_{22}}{N_1}
\right)^2+\frac{g_{12}g_{21}}{N_1N_2} }.
\end{eqnarray}
This is the formula first obtained by Suhl et al.\cite{suh59}
In the case where three bands are equivalent, we can assume simply that
$g_{ii}=g$, $N_i=N(0)$ and $g_{ij}=\nu$ (i,j=1,2,3).  In this case we have
\begin{equation}
\alpha_j= -\Big[ N_j{\rm ln}\left(\frac{2e^{\gamma}\omega_c}{\pi T}\right)-
\frac{g+\nu}{(g+2\eta)(g-\nu)}\Big],
\end{equation}
\begin{equation}
\gamma_{ij}= \frac{\nu}{(g+2\nu)(g-\nu)}.
\end{equation}

Here we investigate the magnetic properties near the upper critical
field $H_{c2}$.  
The Ginzburg-Landau equations are
\begin{equation}  
\alpha_1\psi_1-\gamma_{12}\psi_2-\gamma_{13}\psi_3+\beta_1|\psi_1|^2\psi_1
-K_1\left(\nabla+i\frac{2\pi}{\phi_0}{\bf A}\right)^2\psi_1=0,
\end{equation}
\begin{equation}  
\alpha_2\psi_2-\gamma_{21}\psi_1-\gamma_{23}\psi_3+\beta_2|\psi_2|^2\psi_2
-K_2\left(\nabla+i\frac{2\pi}{\phi_0}{\bf A}\right)^2\psi_2=0,
\end{equation}
\begin{equation}  
\alpha_3\psi_3-\gamma_{31}\psi_1-\gamma_{32}\psi_2+\beta_3|\psi_3|^2\psi_3
-K_3\left(\nabla+i\frac{2\pi}{\phi_0}{\bf A}\right)^2\psi_3=0.
\end{equation}
Let us assume that the magnetic field ${\bf H}$ is along the z axis:
${\bf H}=(0,0,H)$.  We adopt the vector potential ${\bf A}=(0,Hx,0)$.
If we suppose that $\partial\psi_j/\partial y=0$, we obtain for $j$=1,2,3,
\begin{equation}
\alpha_j\psi_j-\sum_{\ell(\neq j)}\gamma_{j\ell}\psi_{\ell}
-K_j\frac{d^2\psi_j}{dx^2}
+ K_j\frac{4\pi^2}{\phi_0^2}H^2x^2\psi_j=0.
\end{equation}
Since this equation is analogous to that for the harmonic oscillator, we can set
\begin{equation}
\psi_j= C_j {\rm exp}\left(-\frac{x^2}{2\xi^2}\right),
\end{equation}
where $C_j$ ($j=1,2,3$) are constants and
\begin{equation}
\xi^2= \frac{\phi_0}{2\pi H}.
\end{equation}
Near the critical point where $H=H_{c2}$, we can neglect the terms
$|\psi_j|^2\psi_j$ since $\psi_j$ are small.
This yields the linearized equation,
\begin{eqnarray}
\left(
\begin{array}{ccc}
\alpha_1+K_1/\xi^2  &  -\gamma_{12}  &  -\gamma_{13}  \\
-\gamma_{21}  &  \alpha_2+K_2/\xi^2  &  -\gamma_{23}  \\
-\gamma_{31}  &  -\gamma_{32}  &  \alpha_3+K_3/\xi^2  \\
\end{array}
\right)
\left(
\begin{array}{c}
C_1  \\
C_2  \\
C_3  \\
\end{array}
\right)=0.\nonumber\\
\end{eqnarray}
To have a nontrivial solution $(C_1,C_2,C_3)\neq 0$,
the secular equation must hold,
\begin{eqnarray}
\left|
\begin{array}{ccc}
\alpha_1+K_1/\xi^2  &  -\gamma_{12}  &  -\gamma_{13}  \\
-\gamma_{21}  &  \alpha_2+K_2/\xi^2  &  -\gamma_{23}  \\
-\gamma_{31}  &  -\gamma_{32}  &  \alpha_3+K_3/\xi^2  \\
\end{array}
\right|=0.
\end{eqnarray}
$T_c$ and
the upper critical field $H_{c2}=\phi_0/(2\pi\xi^2)$ is determined from this
third-order secular equation.
In the two-band case this equation is simply
\begin{eqnarray}
\left|
\begin{array}{cc}
\alpha_1/K_1+1/\xi^2  &  -\gamma_{12}/K_1    \\
-\gamma_{21}/K_2  &  \alpha_2/K_2+1/\xi^2    \\
\end{array}
\right|=0.
\end{eqnarray}
The $H_{c2}$ and the coefficients $(C_1,C_2)$ are immediately derived from this as
\begin{equation}
H_{c2}=\frac{\phi_0}{2\pi}\Biggl[ -\frac{1}{2}\left(\frac{\alpha_1}{K_1}
+\frac{\alpha_2}{K_2}\right)
+\sqrt{ \left(\frac{\alpha_1}{2K_1}-\frac{\alpha_2}{2K_2}\right)^2
+\frac{\gamma_{12}\gamma_{21}}{K_1K_2} }\Biggr],
\end{equation}
\begin{equation}
\frac{C_2}{C_1}= \frac{K_1}{\gamma_{12}}\Biggl[ \frac{1}{2}\left(\frac{\alpha_1}{K_1}
-\frac{\alpha_2}{K_2}\right)
+\sqrt{\left(\frac{\alpha_1}{2K_1}-\frac{\alpha_2}{2K_2}\right)^2
+\frac{\gamma_{12}\gamma_{21}}{K_1K_2} }\Biggr].
\end{equation}
Let us consider a simple case for the three-band model, where we assume that
three bands are equivalent: $\alpha_j=\alpha$, $\gamma_{ij}=J(>0)$, $K_j=K$,
$g_{jj}=g$, $g_{ij}=\eta(>0)$ ($i\neq j$) and $N_j=N(0)$.
We easily obtain
\begin{equation}
H_{c2}= \frac{\phi_0}{2\pi K}(-\alpha+2J)
= \frac{\phi_0}{2\pi K}\left[N(0){\rm ln}\left(\frac{2e^{\gamma}\omega_c}{\pi T}\right)
-\frac{1}{g+2\eta}\right]
\end{equation}
\begin{equation}
T_c= \frac{2e^{\gamma}\omega_c}{\pi}{\rm exp}\left(-\frac{1}{N(0)(g+2\eta)}\right).
\end{equation}
In the $n$-band model, these formulae are generalized to
\begin{equation}
H_{c2}= \frac{\phi_0}{2\pi K}[-\alpha+(n-1)J],
\end{equation}
\begin{equation}
T_c= \frac{2e^{\gamma}\omega_c}{\pi}{\rm exp}\left(-\frac{1}{N(0)(g+(n-1)\eta)}\right).
\end{equation}
Hence, both the upper critical field $H_{c2}$ and the critical temperature $T_c$ 
increase as the number of bands is increased.
The increase of the upper critical field, that is, the reduction of the coherence
length $\xi$, due to the multi-band effect should
be noticed.  
In fact, the high $H_{c2}$ has been reported for the Fe pnictide
SmFeAsO$_{0.85}$F$_{0.15}$\cite{sen08}.

We can show that the three-band system is governed by a single Ginzburg-Landau
parameter $\kappa^*$ near $H_{c2}$.
Thus the conventional three-band model does not yields a type-1.5
superconductor\cite{mos09}.
We define
\begin{equation}
\kappa^*= \sqrt{ \frac{(m^*)^2c^2\beta}{2\pi\hbar^2(e^*)^2} },
\end{equation}
where $e^*=2e$, $m^*$ is the effective mass and $\beta$ is the average of
the coefficients $\beta_j$
($j=1,2,3$).  Then we can show, following the
standard procedure\cite{deG,tin}, that the free energy is given as
\begin{equation}
\frac{F}{V}=\frac{1}{8\pi}\Big[ \langle B_z\rangle^2-
\frac{(H_{c2}-\langle B_z\rangle)^2}{1+\beta_A(2\kappa^{*2}-1)}\Big],
\end{equation}
where $\langle B_z\rangle$ is the integral of $B_z$ over the space and $\beta_A$
is a constant.
This is the same formula as for the single-band superconductor if we substitute
$\kappa$ to $\kappa^*$.  
The magnetization is similarly given by
\begin{equation}
-4\pi M=\frac{H_{c2}-H}{\beta_A(2\kappa^{*2}-1)}.
\end{equation}
The multi-band effect that is contained 
through the parameter $\kappa^*$.
Hence, if there are no some reasons concerning the
symmetry of crystals and superconductivity, the superconductor is regarded as
the type 1 or type 2.

\section{Chirality, Kinks and Double Sine-Gordon Equation}

In Section III, we have shown the model that has a solution with
time-reversal symmetry breaking.
This section is devoted to an investigation of this solution  in three-band
superconductors by using the Ginzburg-Landau theory.
We now consider the phase dynamics of the order parameters.

\subsection{Phase Variables}

We write the order parameters as
\begin{equation}
\psi_j= \rho_j e^{i\theta_j},
\end{equation}
where $\rho_j=|\psi_j|$ is a real quantity.
For simplicity, we assume that the coefficients of the Josephson terms are real:
 $\gamma_{ij}=\gamma_{ji}^*=\gamma_{ji}$.
The free energy density is denoted as $f$, that is, the free energy is given by the
integral of $f$ over the space.  $f$ is written as
\begin{eqnarray}
f&=& \sum_j\alpha_j\rho_j^2+\frac{1}{2}\sum_j\beta_j\rho_j^4
-2\gamma_{12}\rho_1\rho_2\cos(\theta_1-\theta_2)\nonumber\\
&-& 2\gamma_{23}\rho_2\rho_3\cos(\theta_2-\theta_3)
-2\gamma_{31}\rho_3\rho_1\cos(\theta_3-\theta_1)\nonumber\\
&-& \sum_jK_j\rho_je^{-i\theta_j}\left(\nabla+i\frac{2\pi}{\phi_0}{\bf A}\right)^2
(\rho_je^{i\theta_j})
+\frac{1}{8\pi}{\bf H}^2.\nonumber\\
\end{eqnarray}
We focus on the role of phases of the order parameters; we assume that
\begin{equation}
\rho_j=\rho,~~K_j=K,
\end{equation}
and define new phase variables
\begin{equation}
\phi=\theta_1+\theta_2+\theta_3,~~\varphi_1=\theta_1-\theta_2,~~
\varphi_2=\theta_2-\theta_3.
\end{equation}
The free energy density is
\begin{eqnarray}
f&=& 3\alpha\rho^2+\frac{3}{2}\beta\rho^4
-2\rho^2[ \gamma_{12}\cos(\varphi_1)
+ \gamma_{23}\cos(\varphi_2) \nonumber\\
&+& \gamma_{31}\cos(\varphi_1+\varphi_2) ] \nonumber\\
&-& 3K\rho\nabla^2\rho+3K\frac{4\pi^2}{\phi_0^2}\rho^2{\bf A}^2
+\frac{1}{3}K\rho^2(\nabla\phi)^2 \nonumber\\
&+& K\frac{4\pi}{\phi_0}\rho^2{\bf A}\cdot\nabla\phi
+\frac{1}{8\pi}{\bf H}^2\nonumber\\
&+&\frac{1}{3}K\rho^2[(\nabla\varphi_1)^2+(\nabla\varphi_2)^2
+(\nabla(\varphi_1+\varphi_2))^2],
\end{eqnarray}
where $\alpha=(1/3)\sum_j\alpha_j$ and $\beta=(1/3)\sum_j\beta_j$.

The stationary conditions with respect to the fields ${\bf A}$, $\rho$ and $\varphi_j$ 
lead to
\begin{equation}
{\bf j}=\frac{c}{4\pi}{\rm rot}{\bf H}=6Kc\frac{2\pi}{\phi_0}\rho^2\left(\frac{1}{3}
\nabla\phi+\frac{2\pi}{\phi_0}{\bf A}\right),
\end{equation}
\begin{eqnarray}
&&6\alpha\rho+6\beta\rho^3-4\rho[\gamma_{12}\cos(\varphi_1)+\gamma_{23}\cos(\varphi_2)
\nonumber\\
&+&\gamma_{31}\cos(\varphi_{31})]-6K\nabla^2\rho+6K\frac{4\pi^2}{\phi_0^2}H^2x^2\rho
+\frac{2}{3}K\rho(\nabla\phi)^2\nonumber\\
&+&\frac{2}{3}K\rho[(\nabla\varphi_1)^2+(\nabla\varphi_2)^2
+ (\nabla(\varphi_2+\varphi_2))^2]=0,
\end{eqnarray}
\begin{equation}
\gamma_{12}\sin\varphi_1+\gamma_{31}\sin(\varphi_1+\varphi_2)-\frac{1}{3}K
[\nabla^2\varphi_1+\nabla^2(\varphi_1+\varphi_2)]=0,
\end{equation}
\begin{equation}
\gamma_{23}\sin\varphi_1+\gamma_{31}\sin(\varphi_1+\varphi_2)-\frac{1}{3}K
[\nabla^2\varphi_2+\nabla^2(\varphi_1+\varphi_2)]=0.
\end{equation}

\subsection{Phase Potential and Chiral States}

In this section we examine the ground state of the system with the potential
\begin{eqnarray}
V&=& -2\gamma_{12}\rho_1\rho_2\cos(\varphi_1)-2\gamma_{23}\rho_2\rho_3\cos(\varphi_2)
\nonumber\\
&-& 2\gamma_{31}\rho_3\rho_1\cos(\varphi_1+\varphi_2).
\end{eqnarray}
If we set $\varphi_3=\theta_3-\theta_1$, we have 
$\varphi_1+\varphi_2+\varphi_3=0$ (mod$2\pi$).
The minimum of this potential is dependent on the signs of the coefficients 
$\gamma_{ij}\rho_i\rho_j$ of the Josephson terms.
We define $\Gamma_1=-2\gamma_{12}\rho_1\rho_2$, $\Gamma_2=-2\gamma_{23}\rho_2\rho_3$,
and $\Gamma_3=-2\gamma_{31}\rho_3\rho_1$.  The potential is written as
\begin{equation}
V= \Gamma_1\cos(\varphi_1)+\Gamma_2\cos(\varphi_2)+\Gamma_3\cos(\varphi_3). 
\end{equation}
We assume that the absolute values $|\Gamma_i|$ are almost equal in magnitude.
Then there are four cases to be examined as shown in Table I.
When all the $\Gamma_i$ are negative, we have the minimum at
$\varphi_1=\varphi_2=\varphi_3=0$ (Case I).  If we change the sign of $\Gamma_3$,
this produces a frustration effect and $\varphi_i$ take fractional values.
For example, when all the $|\Gamma_i|$ are equal, we have a minimum at
($\varphi_1$,$\varphi_2$,$\varphi_3$)=($\pi/3$,$\pi/3$,$4\pi/3$).  
In this state the order parameters are complex and thus the time reversal symmetry is
broken.
The case IV also exhibits a similar state with fractional values of $\varphi_i$.
If all the $|\Gamma_i|$ are the same, the ground state is at
($\varphi_1$,$\varphi_2$,$\varphi_3$)=($2\pi/3$,$2\pi/3$,$2\pi/3$)
(Fig.1).  

\begin{figure}
\begin{center}
\includegraphics[width=10cm]{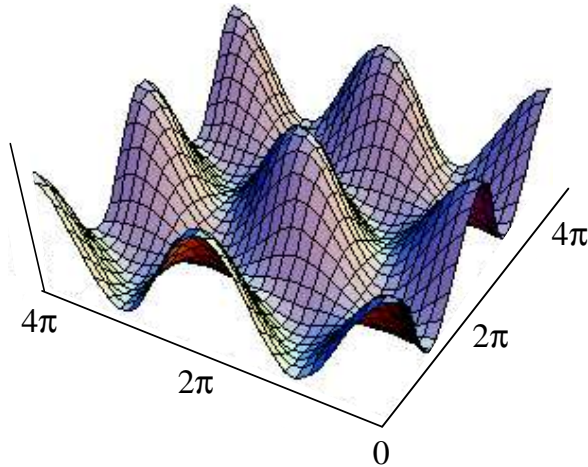}
\caption{
Contour map of $V$ for $\Gamma_1=\Gamma_2=\Gamma_3>0$.
Black and white dots indicate minima of the potential $V$.
Dotted line is the path in the valley connecting two minima.
}
\end{center}
\label{potential}
\end{figure}

\begin{table}
\caption{Classification of the ground state of the potential $V$ of the
Josephson interactions.  $\varphi_i$ in the cases III and IV are for
$|\Gamma_1|=|\Gamma_2|=|\Gamma_3|$.
}
\begin{center}
\begin{tabular}{cccccccc}
\hline
 & $\Gamma_1$ & $\Gamma_2$ & $\Gamma_3$ & $\varphi_1$ & $\varphi_2$ & $\varphi_3$ & \\
\hline
Case I  &  $-$  &  $-$  &  $-$  &  0  &  0  &  0  & \\
Case II &  $+$  &  $+$  &  $-$  &  $\pi$  &  $\pi$  &  0 &  \\
Case III & $-$  &  $-$  &  $+$  &  $\pi/3$  &  $\pi/3$  & $4\pi/3$  & chiral state\\
Case IV &  $+$  &  $+$  &  $+$  &  $2\pi/3$  & $2\pi/3$   & $2\pi/3$  & chiral state\\
\hline
\end{tabular}
\end{center}
\end{table}

\subsection{Double Sine-Gordon Equation and Kinks} 

We consider the solution for $\gamma_{12}=\gamma_{23}$.  In this case
we have a solution with $\varphi_1=\varphi_2\equiv \varphi$.
The variable $\varphi$ satisfies the double sine-Gordon equation,
\begin{equation}
K\nabla^2\varphi-\gamma_{12}\sin\varphi-\gamma_{31}\sin(2\varphi)=0.
\end{equation}
Let us consider the kink solution of the one-dimensional double
sine-Gordon equation.  The energy functional is
\begin{equation}
E= \int \Big[ \frac{1}{2}K_0\left(\frac{d\varphi}{dx}\right)^2
+V(\varphi)\Big]dx,
\end{equation}
where $K_0=2K\rho^2$ and the potential $V$ is
\begin{equation}
V(\varphi)= V_0\left( \cos\varphi+\frac{u}{2}\cos(2\varphi)\right).
\end{equation}
We defined $V_0=-\gamma_{12}\rho^2$ and $u=\gamma_{31}/\gamma_{12}$.
There are two cases to be examined: (1) $\gamma_{12}<0$ and (2) $\gamma_{12}>0$.
We show the classification of the double sine-Gordon model in Table II.

First consider
the case (1) $V_0>0$.
The potential $V(\varphi)$ has a minimum at 
$\varphi=\varphi_0\equiv\cos^{-1}(-1/(2u))$
if $u>1/2$:
\begin{equation}
V(\varphi_0)= V_0\left( -\frac{1}{4u}-\frac{u}{2}\right).
\end{equation}
For $u\leq 1/2$, we have a minimum at $\varphi=\pi$:
\begin{equation}
V(\pi)= V_0\left( -1+\frac{u}{2}\right).
\end{equation}
In the case $u>1/2$ we have a chiral state at $\varphi=\varphi_0$ and a kink solution
that travels from one minimum to the other minimum.
The double sine-Gordon equation
\begin{equation}
\frac{d^2\varphi}{dx^2}= -\frac{V_0}{K_0}(\sin\varphi+u\sin(2\varphi)),
\end{equation}
can be integrated for the boundary conditions
$d\varphi/dx\rightarrow 0$ and $\varphi\rightarrow \varphi_+$ as $x\rightarrow\infty$
and
$d\varphi/dx\rightarrow 0$ and $\varphi\rightarrow -\varphi_+$ as $x\rightarrow -\infty$.
Here $\varphi_+$ is a solution of $\cos(\varphi)=-1/(2u)$ in the range of
$-\pi<\varphi<\pi$.
Since the equation above is integrated as
\begin{equation}
\left(\frac{d\varphi}{dx}\right)^2=\frac{2V_0}{K_0}\left(\cos\varphi+\frac{u}{2}
\cos(2\varphi)+\frac{1+2u^2}{4u}\right),
\end{equation}
we obtain the kink solution as
\begin{equation}
\varphi(x)= \tan^{-1}\left( \frac{ \frac{1}{2u}-q(x)^{-1} }{\sqrt{1-\frac{1}{4u^2}}}
\right)
-\tan^{-1}\left( \frac{ \frac{1}{2u}-q(x) }{\sqrt{1-\frac{1}{4u^2}}}\right),
\end{equation}
where $q(x)$ is defined as
\begin{equation}
q(x)= \exp\left( \sqrt{u^2-\frac{1}{4}}\sqrt{\frac{2V_0}{K_0}}x\right).
\end{equation}
This solution represents the one-kink or one-antikink solution.
Since $\varphi_+/\pi$ takes a fractional value, namely, $\varphi_+=\pm 2\pi/3$ for $u=1$, 
we call this the fractional-$\pi$ kink.
The kink structure is shown in Fig.2.

In the case $V_0>0$ and $u\leq 1/2$, there are minima at $\varphi=\pi ({\rm mod}2\pi)$ in 
the potential $V$.
Thus, we have a $2\pi$-kink solution in this case.
The boundary conditions should be $\varphi\rightarrow\pi$ as $x\rightarrow\infty$
and $\varphi\rightarrow -\pi$ as $x\rightarrow -\infty$, or vice versa.
Since we obtain
\begin{equation}
\left(\frac{d\varphi}{dx}\right)^2=4\frac{V_0}{K_0}(1-2u)\cos^2\left(\frac{\varphi}{2}
\right)\left( 1+\frac{2u}{1-2u}\cos^2\left(\frac{\varphi}{2}\right)\right),
\end{equation}
the kink solution is given by
\begin{equation}
\varphi(x)= \cos^{-1}\left( 1-\frac{2\sinh^2(rx)}{\cosh^2(rx)-2u}\right),
\end{equation}
where $r=\sqrt{V_0(1-2u)/K_0}$. 
 
Second, let us consider the case (2) $V_0<0$.
There are minima at $\varphi=0$ (mod$2\pi$) for $u>-1/2$, and thus the
2$\pi$ kink solution exists.  We obtain
\begin{equation}
\varphi(x)= \cos^{-1}\left( \frac{2\sinh^2(sx)}{\cosh^2(sx)+2u}-1\right),
\end{equation}
where $s=\sqrt{|V_0|(1+2u)/K_0}$.
The kinks in this case are presented in Fig.3.
For large $u$, the kink shows a characteristic at $x=0$ because the
potential has a local minimum at $\varphi=0$ for $u>1/2$.
We have a possibility to find some specific features in the excited state
due to this anomaly.
For $u<-1/2$ we have a fractional-$\pi$ kink that is given by
\begin{equation}
\varphi(x)= \tan^{-1}\left( \frac{1+2|u|t(x)}{\sqrt{4u^2-1}}\right)
-\tan^{-1}\left( \frac{1+2|u|/t(x)}{\sqrt{4u^2-1}}\right),
\end{equation}
where
\begin{equation}
t(x)= \exp\left( \sqrt{2\frac{|V_0|}{K_0}|u|\left(1-\frac{1}{4u^2}\right)}x\right).
\end{equation}
For $u=-1$, this chiral solution satisfies the boundary condition that
$\varphi\rightarrow -\pi/3$ as $x\rightarrow -\infty$ and
$\varphi\rightarrow \pi/3$ as $x\rightarrow \infty$ as shown in Fig.4.

\begin{figure}
\begin{center}
\includegraphics[width=7cm]{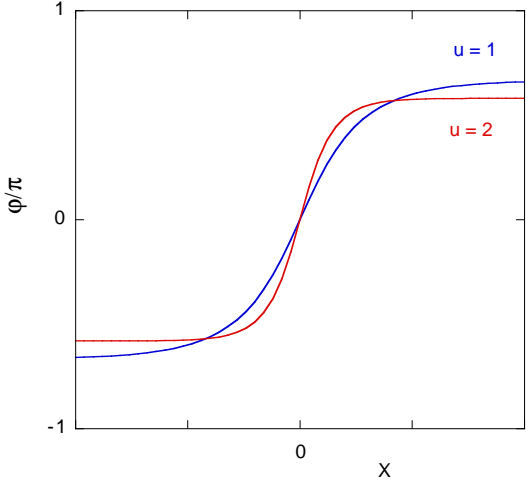}
\caption{
$\varphi$ as a function of $X=\sqrt{2V_0/K_0}x$ for $V_0>0$ and $u=1$, 2.
Fractional kink structure is shown.  For $u=1$, $\varphi(-\infty)=-2\pi/3$ and
$\varphi(\infty)=2\pi/3$ which we call the $4\pi/3$ kink.
}
\end{center}
\label{fkink1}
\end{figure}

\begin{figure}
\begin{center}
\includegraphics[width=7cm]{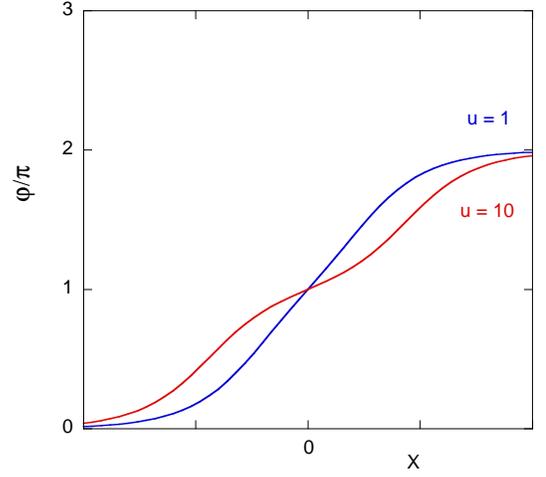}
\caption{
$\varphi$ as a function of $X=\sqrt{(u+1/2)(2|V_0|/K_0)}x$ for $V_0<0$ and $u=1$
and 10.
This shows $2\pi$-kink structure.  For large $u$, the kink shows a saddle-like 
structure at $x=0$.  This is because the potential $V$ has a local minimum
at $\varphi=\pi$ for $u>1/2$.
}
\end{center}
\label{2pkink}
\end{figure}

\begin{figure}
\begin{center}
\includegraphics[width=7cm]{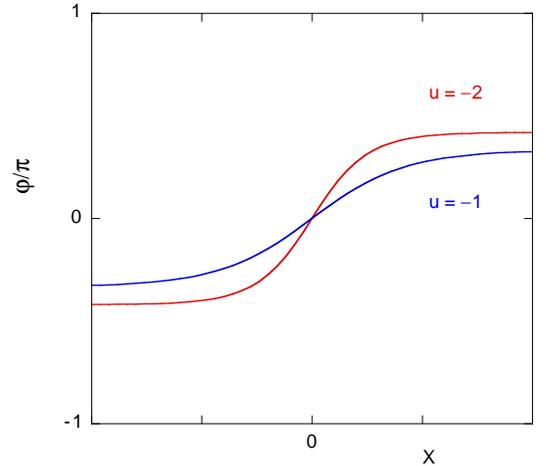}
\caption{
$\varphi$ as a function of $X=\sqrt{2|V_0|/K_0}x$ for $V_0<0$ and $u=-1$ and $-2$.
This shows $2\pi$-kink structure. 
For $u=-1$, $\varphi(-\infty)=-\pi/3$ and $\varphi(\infty)=\pi/3$.
}
\end{center}
\label{fkink2}
\end{figure}

\subsection{Bogomol'nyi-type Inequality}
 
The Bogomol'nyi inequality holds for the double sine-Gordon model with
$V_0>0$ and $u>1/2$ or $V_0<0$ and $u<-1/2$.
Here we call the double sine-Gordon model with this condition the chiral
sine-Gordon model.
The energy functional for the
one-dimensional system $-L/2<x<L/2$ is
\begin{equation}
E= \int_{-L/2}^{L/2} \Big[ \frac{1}{2}K_0\left(\frac{d\varphi}{dx}\right)^2
+V(\varphi)\Big]dx.
\end{equation}
We assume that $L$ is sufficiently large.
If $\varphi$ satisfies the stationary condition $\delta E/\delta\varphi=0$,
we obtain $K_0d^2\varphi/dx^2=dV(\varphi)/d\varphi$.  From this we have
\begin{equation}
\frac{1}{2}K_0\left(\frac{d\varphi}{dx}\right)^2 = V(\varphi)-C_0,
\end{equation}
where $C_0=V(\varphi_0)$  with the condition that 
$d\varphi/dx\rightarrow 0$
as $\varphi\rightarrow \varphi_0$.  Then the energy is
\begin{equation}
E= C_0L+\sqrt{2K_0}\int_{\varphi_-}^{\varphi_+}\sqrt{V(\varphi)-C_0}d\varphi,
\end{equation} 
where $\varphi_+=\varphi(L/2)$ and $\varphi_-=\varphi(-L/2)$, and both
satisfy $d\varphi/dx\rightarrow 0$ as $\varphi\rightarrow\varphi_{\pm}$. 
For $V_0>0$ and $u>1/2$, we obtain the energy of fractional-$\pi$ kink state as
\begin{eqnarray}
E_{f-kink}&=& 2\sqrt{2K_0V_0u}\Big[ \sqrt{1-\frac{1}{4u^2}}
+\frac{1}{2u}\cos^{-1}\left(-\frac{1}{2u}\right)\Big]\nonumber\\
&+& C_0L,
\label{fenergy}
\end{eqnarray}
where $C_0=V(\varphi_0)=V(\varphi_+)=V(\varphi_-)=-V_0(1/(4u)+u/2)$.
This coincides with the energy obtained by substituting the kink solution
directly to the energy functional.
The energy of the 2$\pi$-kink for $V_0<0$ and $u>0$ is
\begin{eqnarray}
E_{2\pi-kink}&=& C_1L+4\sqrt{K_0V_0}\Big[ \sqrt{1+2u}\nonumber\\
&-&\frac{1}{2\sqrt{2u}}
\log\Big|\left(1-\frac{2u}{1+2u}\right)/
\left(1+\frac{2u}{1+2u}\right)\Big|\Big],\nonumber\\
\end{eqnarray}
where $C_1=-|V_0|(1+u/2)$.

Now we derive an inequality for the energy.
Let us consider the case $V_0>0$ and $u>1/2$.
By using the inequality $a^2+b^2\geq 2|ab|$ for real $a$ and $b$, we obtain
\begin{eqnarray}
E&=& \int\Big[ \frac{1}{2}K_0\left(\frac{d\varphi}{dx}\right)^2
+V_0\left( \cos\varphi+\frac{u}{2}\cos(2\varphi)\right)\Big]dx\nonumber\\
&=& C_0L+\int\Big[ \frac{1}{2}K_0\left(\frac{d\varphi}{dx}\right)^2
+V_0 u\left(\cos\varphi+\frac{1}{2u}\right)^2\Big]dx\nonumber\\
&\geq& C_0L+2\sqrt{\frac{1}{2}K_0V_0u}\int\Big|\frac{d\varphi}{dx}\left(
\cos\varphi+\frac{1}{2u}\right)\Big|dx\nonumber\\
&=&\sqrt{2K_0V_0u}\Big[ \sin\varphi_+-\sin\varphi_-+\frac{1}{2u}(\varphi_+
-\varphi_-)\Big]\nonumber\\
&+& C_0L,
\end{eqnarray}
for the one-kink solution that satisfies $d\varphi/dx\geq 0$ and
$\cos\varphi+1/(2u)\geq 0$.
In the case of one fractional-$\pi$ kink shown above, we obtain
\begin{eqnarray}
E&\geq& \sqrt{2K_0V_0u}\Big[ 2\sqrt{1-\frac{1}{4u^2}}+\frac{1}{u}
\cos^{-1}\left(-\frac{1}{2u}\right)\Big]
+C_0L. \nonumber\\
\end{eqnarray}
where we adopt that $0\leq \cos^{-1}\left(-\frac{1}{2u}\right)\leq\pi$.
The lower bound of the energy coincides with the energy in eq.(\ref{fenergy}).
Here, we define the conserved current
\begin{equation}
J^{\mu}= \frac{1}{2A}\epsilon^{\mu\nu}\partial_{\nu}\varphi,
\end{equation}
with the charge
\begin{equation}
Q=\int_{-\infty}^{\infty}J^0(x)dx=\frac{1}{2A}[\varphi(x=\infty)
-\varphi(x=-\infty)],
\end{equation} 
where $\epsilon^{\mu\nu}$ is the antisymmetric symbol and $x^0=t$, $x^1=x$.
$A$ is the normalization constant defined by $A=\cos^{-1}(-1/(2u))$ with the value
in the range $0\leq A\leq\pi$.
$\partial_{\mu}J^{\mu}=0$ follows immediately from the antisymmetric symbol
$\epsilon^{\mu\nu}$ and does not depend on the equation of motion.
The kink has $Q=1$, and an antikink with $Q=-1$ exists that has a configuration
with $\varphi(-\infty)=A$ and $\varphi(\infty)=-A$.
If we assume that $\cos\varphi+1/(2u)>0$ for the kink solution, 
the energy inequality is written as
\begin{eqnarray}
E&\geq& {\rm sign}Q\sqrt{2K_0V_0u}(\sin\varphi_+-\sin\varphi_-)
+\sqrt{\frac{2K_0V_0}{u}}A|Q|\nonumber\\
&+&C_0L.
\end{eqnarray}
This is an inequality of Bogomol'nyi type.
For the kink satisfying $\cos\varphi+1/(2u)<0$, the inequality is
\begin{eqnarray}
E&\geq& -\Big[{\rm sign}Q\sqrt{2K_0V_0u}(\sin\varphi_+-\sin\varphi_-)\nonumber\\
&+& \sqrt{\frac{2K_0V_0}{u}}A|Q|\Big]
+ C_0L.
\end{eqnarray}

Since $\varphi(\pm\infty)$ can take values 
$\varphi(\infty)=\pm A+2n_{+}\pi$ and $\varphi(-\infty)=\pm A+2n_{-}\pi$ with
integers $n_+$ and $n_-$,
there exists a wide spectrum of solitons
such that $Q=\pm 1+(n_+-n_-)\pi/A$ or $Q=(n_+-n_-)\pi/A$.
In general, the $n$-kink solution is a sum of (anti)kinks with plateaus where 
$d\varphi/dx$ vanishes.
The total kink energy is a summation of contributions from each (anti)kink structure.
Suppose that $d\varphi/dx=0$ at $x=x_1,\cdots,n-1$ in an $n$-kink solution.
We set $x_0=\infty$ and $x_n=-\infty$.  We define the charge $Q_j$ for each kink
component: $Q_j=\varphi(x_{j-1})-\varphi(x_j)$.  
$Q=\sum_jQ_j$ holds.
The energy bound is
\begin{eqnarray}
E&\geq& \sqrt{2K_0V_0u}\sum_js_j{\rm sign}Q_j(\sin\varphi(x_{j-1})-\sin\varphi(x_{j}))
\nonumber\\
&+& \sqrt{\frac{2K_0V_0}{u}}\frac{1}{2}\tilde{Q} +C_0L,
\end{eqnarray}
where $s_j=1$ if $\cos\varphi+1/(2u)>0$ in the $j$-th (anti)kink and $s_j=-1$ if 
$\cos\varphi+1/(2u)<0$, and we defined
\begin{equation}
\tilde{Q}=\sum_js_j|Q_j|.
\end{equation}

\begin{table}
\caption{Classification of the double sine-Gordon model with the potential
$V(\varphi)=V_0(\cos(\varphi)+(u/2)\cos(2\varphi))$.
$V(\varphi)$ has minima at $\varphi=\varphi_0$ (mod$2\pi$).
}
\begin{center}
\begin{tabular}{ccccc}
\hline
 $V_0$ & $u$ & $\varphi_0$ & kink  &  \\
\hline
$V_0>0$  &  $u>1/2$  &  $\cos^{-1}(-1/(2u))$  &  fractional-$\pi$ kink &  chiral \\
$V_0>0$  &  $u<1/2$  &  $\pi$  &  2$\pi$-kink  &   \\
$V_0<0$  &  $u> -1/2$  &  $0$  &  2$\pi$-kink  &   \\
$V_0<0$  &  $u<-1/2$  &  $\cos^{-1}(-1/(2u))$  & fractional-$\pi$ kink &  chiral \\
\hline
\end{tabular}
\end{center}
\end{table}

\section{Fractional Vortices and Bound States}

In general, there are solutions of vortices with fractional quantum flux in
multi-band superconductors.  Kinks in the space of phase variables $\theta_j$
play a central role for the existence of fractional flux vortices.
In the two-band model, the half-quantum-flux vortex exists with a line of
singularity of the phase variables $\theta_j$ as shown in Fig.5.
Here $\theta_1$ changes from 0 to $\pi$ (or $\pi$ to 0) across the cut, and 
simultaneously
$\theta_2$ changes from 0 to $-\pi$ (or $-\pi$ to 0).
In the case of Fig.5, a net-change of $\theta_1$ is $2\pi$ by
a counterclockwise encirclement of the vortex, and that of $\theta_2$
vanishes, due to singularities.
Thus, we have a half-quantum flux vortex.
This is an interpretation of half-quantum flux vortices in triplet 
superconductors\cite{kee00} in terms of the phase of the order parameter.

\begin{figure}
\begin{center}
\includegraphics[width=10cm]{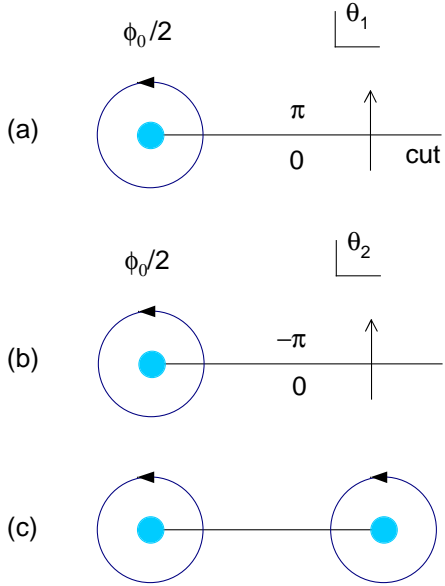}
\caption{
Half-quantum flux vortex with line singularity.
The phase variables $\theta_1$ and $\theta_2$ have line singularities,
as shown in (a) and (b).
Two half-flux vortices connected by the singularity are shown in (c).
}
\end{center}
\label{2-vortex}
\end{figure}

In three-band superconductors, the fractional-flux vortex exists in the chiral
case as well as the non-chiral case.
Since we have the fractional-$\pi$ kink in the chiral state (cases III and IV), 
the new types of vortices with fractional
flux quanta exist on a domain wall of the kink\cite{tan10a}.
The kink considered in the previous section is a one-dimensional structure in 
superconductors.  
There are many types of kinks connecting two minima of the potential in three-band
superconductors.

\begin{figure}
\begin{center}
\includegraphics[width=10cm]{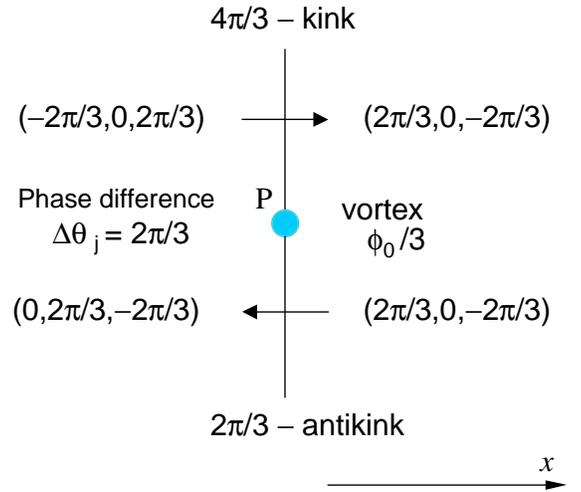}
\caption{
Kink, antikink and a fractional flux vortex for $\Gamma_1=\Gamma_2=\Gamma_3>0$.
The vortex is at the point $P$ with flux $\phi_0/3$ where
$\phi_0$ is the flux quantum.
We start from $(\theta_1,\theta_2,\theta_3)=(-2\pi/3,0,2\pi/3)$ to reach
$(0,2\pi/3,-2\pi/3)$ (modulo $2\pi$) through the $4\pi/3$-kink and $2\pi/3$-antikink.
$\varphi_1=\theta_1-\theta_2$ goes from $-2\pi/3$ to $2\pi/3$ crossing the
$4\pi/3$-kink, and $\varphi_1$ goes from $2\pi/3$ to $4\pi/3\equiv -2\pi/3$ (mod $2\pi$)
through the $2\pi/3$-kink. 
}
\end{center}
\label{fvortex}
\end{figure}

\begin{figure}
\begin{center}
\includegraphics[width=8cm]{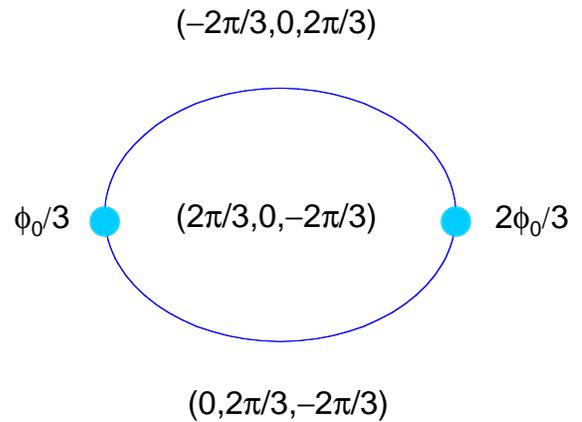}
\caption{
Two-vortex bound state with line singularities in the time-reversal
symmetry broken state.
The phase variables $\theta_i$ ($i=1,2,3$) have singularities that are described by
kinks in Fig.6.
The total flux is $\phi_0$.  Topologically, the flux $2\phi_0/3$ is equivalent to
$-\phi_0/3$.  Thus, this state corresponds to the meson under the duality
transformation between charge and magnetic flux.
}
\end{center}
\label{2-fvortex}
\end{figure}

Let us discuss the fractional vortices in the three-band model here.
Suppose that two kinks, one is a kink and the other is an antikink,
intersect at a point $P$ as shown in Fig.6 in a two-dimensional $xy$-plane.
If a vortex exists along the $z$ axis just at the point $P$, the vortex should
have a fractional flux quantum so that the phase change around the point $P$ is
$2\pi$.  For $V_0>0$ and $u>1/2$, this is shown schematically in Fig.6.
We set the phases of the order parameters $(\theta_1,\theta_2,\theta_3)=(-2\pi/3,0,2\pi/3)$
in some region. After crossing the $4\pi/3$ kink, they become $(2\pi/3,0,-2\pi/3)$
where the phase variables $\varphi_1$ and $\varphi_2$ change from $-2\pi/3$ to
$2\pi/3$.  If there is also a domain wall of an antikink that starts from the point
$P$ as in Fig.6, we have the phases $(0,2\pi/3,-2\pi/3)$ after we cross
the antikink.  Here, we obtain the phase difference between the initial and final
states (see Fig.6).  In this case, the vortex that is located through  the 
point $P$ along the $z$ axis should have a fractional flux quantum $\phi_0/3$.  Thus,
in the chiral region of three-band superconductors,  the existence of fractional vortices
is easily concluded in this way.

\begin{figure}
\begin{center}
\includegraphics[width=10cm]{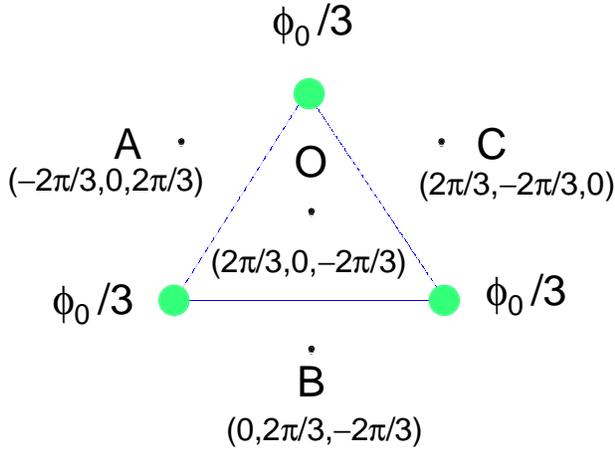}
\caption{
Three-vortex bound state with line singularities in the time-reversal
symmetry broken state.
Each vortex has $\phi_0/3$ and the total flux is $\phi_0$.
The phase variables $\theta_i$ ($i=1,2,3$) have singularities that are
fractional-$\pi$ kinks.
}
\end{center}
\label{3-vortex1}
\end{figure}

In the three-band model, the fractional vortex has two line singularities (kinks) in
the phases of the gap function as shown in Fig.6.
From Fig.6, we have a two-vortex bound state as presented in 
Fig.7 in the chiral state.  
Two vortices form a 'molecule' by two kinks.
This state may have lower energy than the vortex state with quantum flux $\phi_0$ since
the magnetic energy $(5/9)\phi_0^2$ is smaller than $\phi_0^2$ of the unit flux.
The energy of kinks is proportional to the distance $R$ between two fractional
vortices if $R$ is large.  Thus, the attractive interaction works between them
if $R$ is sufficiently large.

Three-vortex bound states are also formulated:  they are shown in Figs.8, 9 and 10.
The first two figures indicate bound states in the time-reversal symmetry broken 
state.
The last one is for the unbroken state\cite{nit10}.
These states correspond to baryons if we regard the magnetic flux as charge.

\begin{figure}
\begin{center}
\includegraphics[width=10cm]{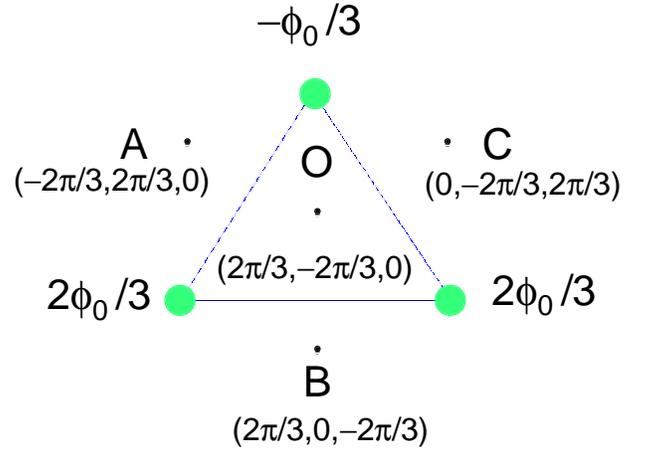}
\caption{
Three-vortex bound state with line singularities in the time-reversal
symmetry broken state.
This state corresponds to the proton.
}
\end{center}
\label{3-vortex2}
\end{figure}

\begin{figure}
\begin{center}
\includegraphics[width=10cm]{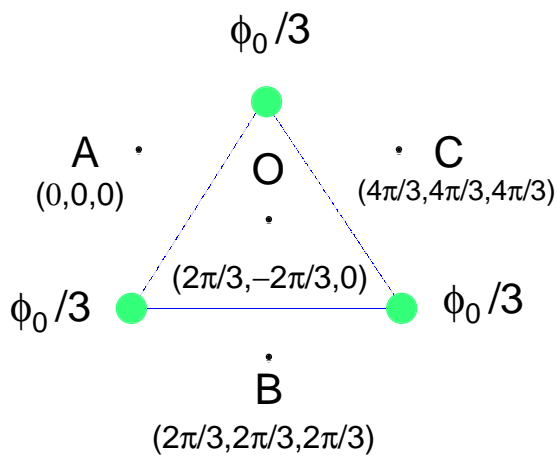}
\caption{
Three-vortex bound state with line singularities in the time-reversal
symmetric state.
Each vortex has $\phi_0/3$ and the total flux is $\phi_0$.
In this state, the region including the point O has higher energy.
}
\end{center}
\label{3-vortex3}
\end{figure}

\section{Applications}

In this Section, we briefly consider some applications.
In the one-band model, the symmetry of Cooper pairs is classified into irreducible
representations and each representation is one dimensional in most cases except
the two-dimensional representation for $p$-wave symmetry\cite{hlu99,kon01,yan08}.
In the one-dimensional
representation, the phase of the order parameter is not important, except the case
where two states in two different representations accidentally are degenerate.
In multi-band systems, the relative phase between bands becomes important and
plays an essential role.

A promising candidate of a multi-band superconductor
is a superconductor-insulator-superconductor
(SIS) junction.  An SIS tri-junction will be a three-band superconductor,
interacting through weak Josephson couplings,
with three equivalent bands.
We hope that SIS junctions or superlattice systems  will be realized
as artificial multi-band systems.

The Fe pnictides in general contain several bands and it is probable that the order
parameters in these bands belong to the same irreducible representation.
We can expect high $H_{c2}$ due to the multiplicity of bands in Fe pnictides.
If we assume that the pairing interactions are frustrating, that is, in the cases
III and IV, or $V_0>0$ in the potential $V$, there is a possibility that chiral 
superconductivity 
will emerge in Fe pnictides superconductors.
Let us examine the pairing interactions that originate from the spin susceptibility
$\chi({\bf q})$.  It has been shown that $\chi({\bf q})$ has peaks at ${\bf q}=(\pi,0)$
and $(0,\pi)$ in multi-band models for Fe pnictides\cite{maz08,kur08}.
In some Fe pnictides $\chi({\bf q})$ has a peak at ${\bf q}=(\pi.\pi)$ as well as
at ${\bf q}=(\pi,0)$\cite{kur08}.
In this case we have a frustration between the pairing interaction due to
$\chi(\pi,0)$ and $\chi(\pi,\pi)$.  This may lead to a chiral $s$-state $s_{chiral}$
rather than $s_{\pm}$ and $s_{++}$ as an intermediate state of these two states.

\section{Summary and Discussion}

In this paper we have investigated multi-band superconductors.
We have examined time-reversal symmetry breaking and fractional flux vortices in
the three-band model.
Our discussion is based on the mechanism of time-reversal symmetry breaking
due to the degeneracy of the gap equation.
The other mechanism should also be investigated in future studies\cite{imr75}.
We derived the Ginzburg-Landau free energy for multi-component superconductors
with more than two components.  The coefficient of each term is expressed in terms
of the inverse of the matrix of pairing interactions $G=(g_{ij})$.  
We have shown that the system is determined by the single Ginzburg-Landau
parameter $\kappa^*$ near the upper critical field $H_{c2}$.
For the frustrating pairing interactions, the chiral superconducting
state exists with broken time reversal symmetry. 
The double sine-Gordon model appears as an effective theory to describe low-excitation
energy states for three-component superconductors.
In the chiral case, this model has solutions of fractional-$\pi$ kinks and satisfies
the inequality of Bogomol'nyi type with the topological charge $Q$.
We also discussed the existence of fractional flux vortices and that of the
multi-vortex bound state in multi-band
superconductors.

The kinks form domains, which we call the chiral domains, in multi-band superconductors; 
this is analogous to the domains in triplet 
superconductors\cite{sig99}.
In the chiral region of three-band superconductors, there exist several types of
kink solitons.
Vortices with fractional flux quanta exist on domain walls.
It is important to investigate the effect of domain walls on the vortex dynamics.
Domain walls play an important role in strong pinning of vortices and will be closely
related to flux-flow noise.
The generation of chiral domains by quenching superconductors is also an attractive
subject.  This is a phenomenon caused by the Kibble mechanism in 
superconductors\cite{kib76,zur85}.
In experiments, the number of domains may be controlled by quenching speed and
sample morphology that pins domain walls.

This work was supported by a Grant-in Aid for Scientific Research from the
Ministry of Education, Culture, Sports, Science and Technology of Japan.

\end{document}